\documentclass[%
 reprint,
 amsmath,amssymb,
 aps,
]{revtex4-1}

\usepackage{graphicx}
\usepackage{dcolumn}
\usepackage{bm}
\usepackage{multirow}

\begin{document}

\preprint{APS/123-QED}

\title{Confinement potentials for the study of heavy mesons}

\author{Alfredo Vega}
\email{alfredo.vega@uv.cl}
 
\author{Francisca Rojas}%
\email{francisca.rojasgo@alumnos.uv.cl}
\affiliation{%
 Instituto de F\'isica y Astronom\'ia, \\
 Universidad de Valpara\'iso,\\
 A. Gran Breta\~na 1111, Valpara\'iso, Chile
}%

\date{\today}

\begin{abstract}


We consider a Coulomb potential plus a confinement potential $ A r^{\nu} $ and we study which of the two terms is dominant in the description of quarkonia. We find that in general the term of confinement is dominant, which allows us to understand why such potentials, like Martin's potential, are successful in describing of heavy mesons.

\begin{description}
\item[PACS numbers]

\end{description}
\end{abstract}

\pacs{Valid PACS appear here}
\maketitle


\section{\label{sec:level1}Introduction}

Currently, there is experimental evidence that hadrons are bound states of quarks and gluons, whose interactions are described through quantum chromodynamics (QCD). This quantum field theory has a coupling constant that shows both a perturbative and a non-perturbative region, and it is precisely in the latter where several properties of interest in Hadronic physics are found (which makes its use difficult). This situation, together with motivating the development of techniques and tools that allow the direct use of QCD in the study of hadrons such as Lattice QCD (e.g.[1]) or the use of the Dyson Schwinger equations to study hadrons (e.g.[2]), has also prompted the development of phenomenological models, which, inspired by QCD, capture important properties of the interaction between quarks and gluons, and can thus perform calculations of hadronic properties. 
Potential models between quarks [3-7], which have remained valid since the mid-1970s, when the $J/\psi$ was discovered (the first known hadron with heavy quarks), when are applied to quarkonia, good results can be obtained from the use of the Schr\"odinger equation together with a potential that describes the interaction between the heavy quarks that make up the said hadron.


Based on QCD we can only have an idea of the behavior of the potential when the quarks are very close and far away. When the quarks are close, we are in the region where $q^{2}$ is large, and therefore the coupling constant is small, which allows us to use perturbation theory to study the interaction between quarks considering only the one gluon exchange, which allows a Coulomb potential to be extracted. At the other extreme, we have that at low $q^{2}$ the coupling constant becomes large, which prevents the use of perturbation theory, and we can only say that in this region the quarks are confined, but we have no a way to access this part of the potential precisely (although Lattice suggests a linear term). Thus we know that a potential between quarks well motivated by QCD must interpolate between a Coulomb potential at short distances and confinement potential at great distances, for which several alternatives have been proposed (some examples of this can be reviewed in \cite{Lucha:1991vn, Bykov:1985it, Dib:2012vw}), with the simplest (and that in general has yielded very good results) being the one that considers the sum of both contributions.


As we just mentioned in the previous paragraph, a potential well motivated by QCD to describe the interaction between quarks is of the Coulomb form plus a confinement potential, where a habitual choice for this last term considers some power of the distance between the quarks; nevertheless many authors have considered only the confinement part, and with power-law potential have obtained good results in the study of quarkonia \cite{Martin:1980jx, Martin:1980rm}. This type of potential in the literature is usually referred to as merely phenomenological without being well motivated by QCD.


In this paper, we are interested in studying whether potential between quarks that consider only connement should be considered as merely phenomenological as they have been cataloged until now, or if they can be considered the dominant part of well-motivated potentials from QCD. For this we consider the "Coulomb plus power potential ($CPP\nu$)" used in \cite{Rai:2008sc, Patel:2008na}, which is of the form $V(r) = -\alpha/r + A r^{\nu} + V_{0}$; therefore the potential used corresponds to a generalization of the Cornell potential \cite{Eichten:1974af, Eichten:1978tg, Eichten:1979ms}, which considers a linear type confinement. For different values of $\nu$ we will calculate the spectrum and decay constants of the charmonium, and analyze if these models allow us to obtain good results by neglecting the Coulomb part with and without readjusting the parameters. On the other hand, we compare the average radius of the $1S$ states for the full potential with the distance at which the confinement term is the dominant one, thereby extending to several values of $\nu$ the criterion used in \cite{Choudhury:2013kta, Choudhury:2014mva} to decide if the linear term can be considered a disturbance in the Cornell potential.


Before concluding this introduction it is important to mention a potential such as the one considered, which has been associated with the exchange contribution of one gluon and another that captures confinement effects, corresponds to the static limit of the interaction between quarks, and that is why in a more detailed description of quarkonia through the use of potentials should consider relativistic contributions (e.g. see \cite{Lucha:1991vn, Bykov:1985it, Dib:2012vw}), some of which are dependent on angular orbital momentum and spin. Although they give small contributions, they allow to better describe the spectrum, affording a more detailed decription of the observed spectrum, and making differences between states $c\bar{c}$ scalar ($\eta$) and vector ($J/\psi$).


The work is structured as follow: In section II we present the potential model used, we discuss the way in which the parameters have been determined and we present the results obtained for the masses and decay constants for the base state and the first three radial excitations of the charmonium. Section III is dedicated to analyzing if the Coulomb part is greater than the confinement in full potential. Finally in section IV we discuss the results and analyze the values of $\nu$ for which we obtain masses and decay constants close to the measures, and thus shed light on the possible values of $\nu$ in the confinement potential, and if the latter alone can be considered a good approximation for the full potential is motivated by QCD or simply, as has been presented in the literature, it is a purely phenomenological model.





\section{\label{sec:level1}Non-relativistic treatment for systems $Q \bar{Q}$}

As we explained in the introduction, the potential model we will use is the $CPP\nu$ \cite{Rai:2008sc,Patel:2008na}, which we summarize below for the study of the charmonium.


We consider a Hamiltonian given by

\begin{equation}
\label{Hamiltoniano}
H = M + \frac{p^{2}}{2 m} + V(r),
\end{equation}
where
\begin{equation}
M = m_{Q} + m_{\bar{Q}}~~~y~~~m = \frac{m_{Q} m_{\bar{Q}}}{m_{Q} + m_{\bar{Q}}},
\end{equation}
where $ m_ {Q} $ and $ m _ {\ bar {Q}} $ correspond to the masses of the quarks and antiquarks respectively, whose value throughout this work is $ m_ {c} = 1.31 ~ GeV $; $ p $ is the relative momentum of the quarks and $V (r)$ is a potential quark - antiquark of the form
\begin{equation}
\label{Potencial}
V(r) = - \frac{\alpha_{c}}{r} + A r^{\nu} - V_{0}.
\end{equation}

Here $\alpha_{c} = \frac{4}{3} \alpha_{s}$, where $\alpha_{s}$ is the strong coupling constant. In this paper we use $\alpha_{c}(c\bar{c}) = 0.4$; $\nu$ is an exponent that we set in advance, which we have restricted to values between $0.1$ and $2.5$; and finally $A$ is a parameter of the potential that varies for each value of $\nu$ and is adjusted to produce a good fit to the values of the masses of charmonium. In \cite{Rai:2008sc,Patel:2008na} the authors consider an expression obtained using the usual variational method, but in this work we have decided to use the MATHEMATICA program developed in \cite{Lucha:1998xc} to solve the Schr\"odinger equation and with this find values of $A$ that reproduce values close to the masses of $J/\psi$ and radial excitations considered in this work.


We have considered 6 possible families for the parameters $A$ and $V_{0}$. The first three families of parameters examine the case $V_{0}=0$, with $A$ being the only parameter adjusted. For this case, which is summarized in Table I, ($CC$) include the full potential, i.e., the Coulomb part plus the confinement term; in the case ($OC$) the Coulomb part has been discarded without modifying the confinement part, i.e., the value of $A$ obtained in column $CC$ is maintained; and finally the case ($OCR$) examine only confinement again, but this time the $A$ parameter has been determined again for this potential. In each case, the error in the calculation of masses obtained with these parameters is presented. Table II shows the value of the model parameters for the remaining three families, where the same cases as Table I are include, but here $V_{0}$ is determined together with $A$.


Figures I and II show the results of the masses and decay constants of the first four radial excitations of the charmonium for each value of $\nu$ considered.


\begin{table*}
\begin{center}
\begin{tabular}{| c | c | c | c | c | c | c |  c | c | c |}
\hline
$\nu$ & \multicolumn{3}{ c |}{$CC$} & \multicolumn{3}{ c |}{$OC$} & \multicolumn{3}{ c |}{$OCR$} \\ 
\cline{2-10}
& $A$ & $V_{0}$ & Error & $A$ & $V_{0}$ & Error & $A$ & $V_{0}$ & Error \\
\hline
\hline
~~0.1~~ & ~~0.881~~ & ~~0~~ & ~~0.149~~ & ~~0.881~~ & ~~0~~ & ~~0.168~~ & ~~0.817~~ & ~~0~~ & ~~0.166~~ \\ 
 \hline
~~0.2~~ & ~~0.733~~ & ~~0~~ & ~~0.121~~ & ~~0.733~~ & ~~0~~ & ~~0.149~~ & ~~0.666~~ & ~~0~~ & ~~0.144~~ \\ 
 \hline
~~0.3~~ & ~~0.614~~ & ~~0~~ & ~~0.098~~ & ~~0.614~~ & ~~0~~ & ~~0.131~~ & ~~0.550~~ & ~~0~~ & ~~0.124~~ \\ 
 \hline
~~0.4~~ & ~~0.514~~ & ~~0~~ & ~~0.076~~ & ~~0.514~~ & ~~0~~ & ~~0.115~~ & ~~0.456~~ & ~~0~~ & ~~0.106~~ \\ 
 \hline
~~0.5~~ & ~~0.430~~ & ~~0~~ & ~~0.057~~ & ~~0.430~~ & ~~0~~ & ~~0.101~~ & ~~0.378~~ & ~~0~~ & ~~0.089~~ \\
 \hline
~~0.6~~ & ~~0.360~~ & ~~0~~ & ~~0.040~~ & ~~0.360~~ & ~~0~~ & ~~0.089~~ & ~~0.314~~ & ~~0~~ & ~~0.075~~ \\ 
 \hline
~~0.7~~ & ~~0.301~~ & ~~0~~ & ~~0.025~~ & ~~0.301~~ & ~~0~~ & ~~0.079~~ & ~~0.261~~ & ~~0~~ & ~~0.061~~ \\
 \hline
~~0.8~~ & ~~0.251~~ & ~~0~~ & ~~0.013~~ & ~~0.251~~ & ~~0~~ & ~~0.070~~ & ~~0.216~~ & ~~0~~ & ~~0.049~~ \\ 
 \hline
~~0.9~~ & ~~0.209~~ & ~~0~~ & ~~0.008~~ & ~~0.209~~ & ~~0~~ & ~~0.063~~ & ~~0.179~~ & ~~0~~ & ~~0.039~~ \\
 \hline
~~1.0~~ & ~~0.174~~ & ~~0~~ & ~~0.015~~ & ~~0.174~~ & ~~0~~ & ~~0.058~~ & ~~0.148~~ & ~~0~~ & ~~0.030~~ \\ 
 \hline
~~1.1~~ & ~~0.145~~ & ~~0~~ & ~~0.024~~ & ~~0.145~~ & ~~0~~ & ~~0.055~~ & ~~0.123~~ & ~~0~~ & ~~0.022~~ \\
 \hline
~~1.2~~ & ~~0.120~~ & ~~0~~ & ~~0.033~~ & ~~0.120~~ & ~~0~~ & ~~0.052~~ & ~~0.102~~ & ~~0~~ & ~~0.018~~ \\ 
 \hline
~~1.3~~ & ~~0.100~~ & ~~0~~ & ~~0.041~~ & ~~0.100~~ & ~~0~~ & ~~0.052~~ & ~~0.084~~ & ~~0~~ & ~~0.016~~ \\
 \hline
~~1.4~~ & ~~0.083~~ & ~~0~~ & ~~0.048~~ & ~~0.083~~ & ~~0~~ & ~~0.053~~ & ~~0.069~~ & ~~0~~ & ~~0.018~~ \\ 
 \hline
~~1.5~~ & ~~0.069~~ & ~~0~~ & ~~0.055~~ & ~~0.069~~ & ~~0~~ & ~~0.054~~ & ~~0.057~~ & ~~0~~ & ~~0.021~~ \\
 \hline
~~1.6~~ & ~~0.057~~ & ~~0~~ & ~~0.061~~ & ~~0.057~~ & ~~0~~ & ~~0.055~~ & ~~0.047~~ & ~~0~~ & ~~0.026~~ \\ 
 \hline
~~1.7~~ & ~~0.047~~ & ~~0~~ & ~~0.067~~ & ~~0.047~~ & ~~0~~ & ~~0.056~~ & ~~0.039~~ & ~~0~~ & ~~0.030~~ \\
 \hline
~~1.8~~ & ~~0.039~~ & ~~0~~ & ~~0.072~~ & ~~0.039~~ & ~~0~~ & ~~0.058~~ & ~~0.032~~ & ~~0~~ & ~~0.035~~ \\ 
 \hline
~~1.9~~ & ~~0.033~~ & ~~0~~ & ~~0.077~~ & ~~0.033~~ & ~~0~~ & ~~0.065~~ & ~~0.027~~ & ~~0~~ & ~~0.039~~ \\
 \hline
~~2.0~~ & ~~0.027~~ & ~~0~~ & ~~0.082~~ & ~~0.027~~ & ~~0~~ & ~~0.065~~ & ~~0.022~~ & ~~0~~ & ~~0.043~~ \\ 
 \hline
~~2.1~~ & ~~0.022~~ & ~~0~~ & ~~0.086~~ & ~~0.022~~ & ~~0~~ & ~~0.065~~ & ~~0.018~~ & ~~0~~ & ~~0.047~~ \\
 \hline
~~2.2~~ & ~~0.019~~ & ~~0~~ & ~~0.090~~ & ~~0.019~~ & ~~0~~ & ~~0.074~~ & ~~0.015~~ & ~~0~~ & ~~0.051~~ \\ 
 \hline
~~2.3~~ & ~~0.015~~ & ~~0~~ & ~~0.094~~ & ~~0.015~~ & ~~0~~ & ~~0.069~~ & ~~0.012~~ & ~~0~~ & ~~0.054~~ \\
 \hline
~~2.4~~ & ~~0.013~~ & ~~0~~ & ~~0.097~~ & ~~0.013~~ & ~~0~~ & ~~0.078~~ & ~~0.010~~ & ~~0~~ & ~~0.057~~ \\ 
 \hline
~~2.5~~ & ~~0.011~~ & ~~0~~ & ~~0.101~~ & ~~0.011~~ & ~~0~~ & ~~0.083~~ & ~~0.008~~ & ~~0~~ & ~~0.061~~ \\
 \hline
 \hline
\end{tabular}
\caption{Parameters of the potential for the C
charmonium next to the corresponding rms error obtained for the masses of the four charmonia considered in this work for the case where $V_{0} = 0$. The $CC$ and $OC$ indices consider that said parameters were obtained using a Coulomb potential plus confinement ($CC$), confinement only ($OC$) or confinement only a with readjustment of parameters ($OCR$).}
\label{tabla1}
\end{center}
\end{table*}

\begin{center}
\begin{figure*}
  \begin{tabular}{c c c}
    \includegraphics[width=2.3 in]{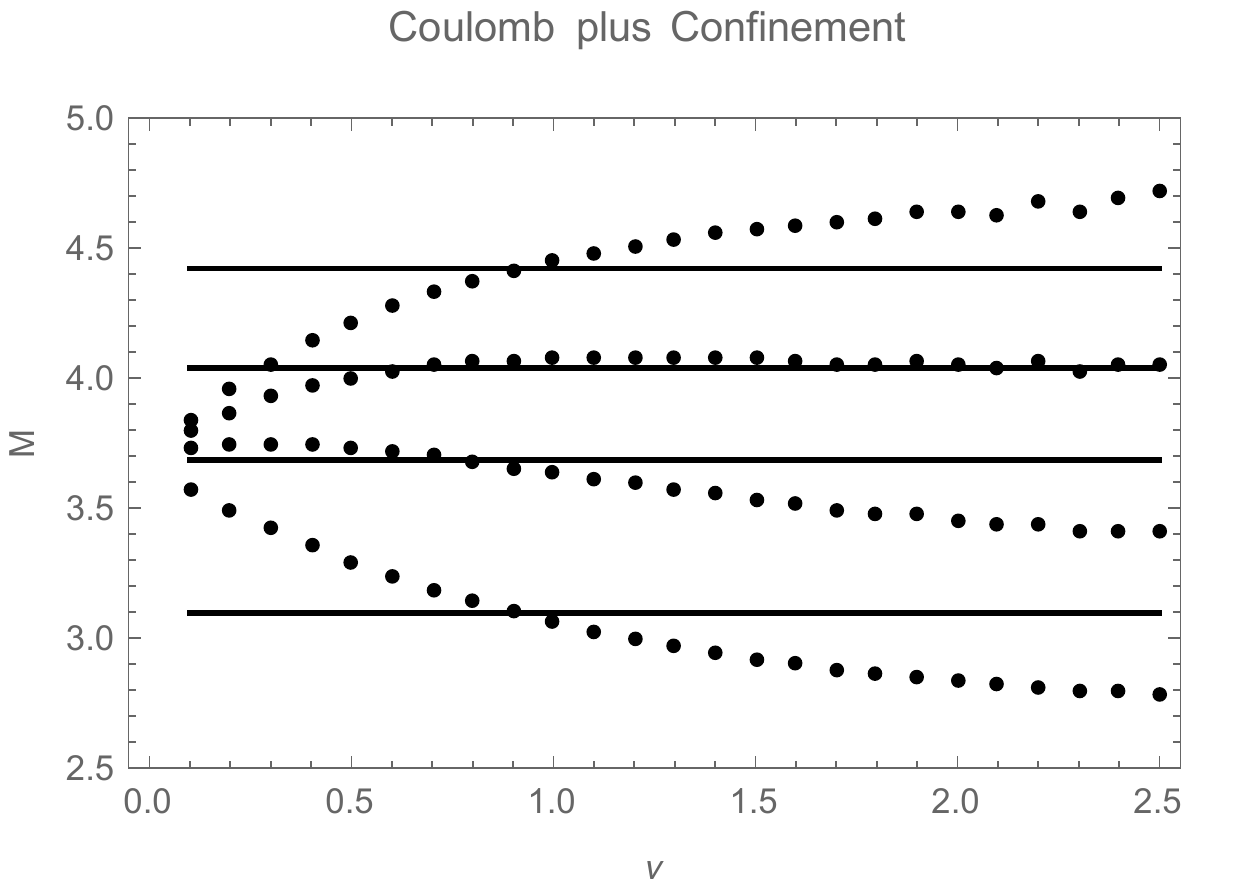}
    \includegraphics[width=2.3 in]{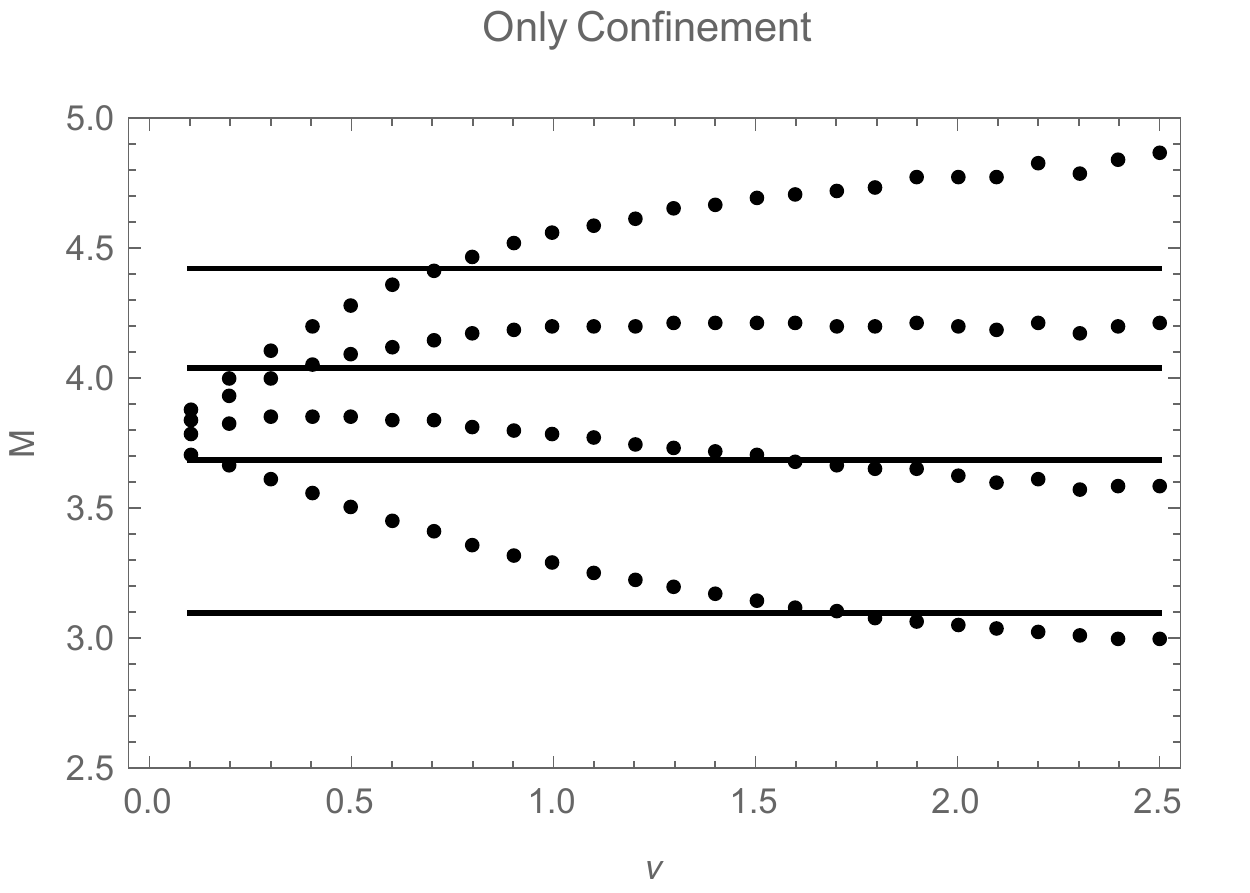}
    \includegraphics[width=2.3 in]{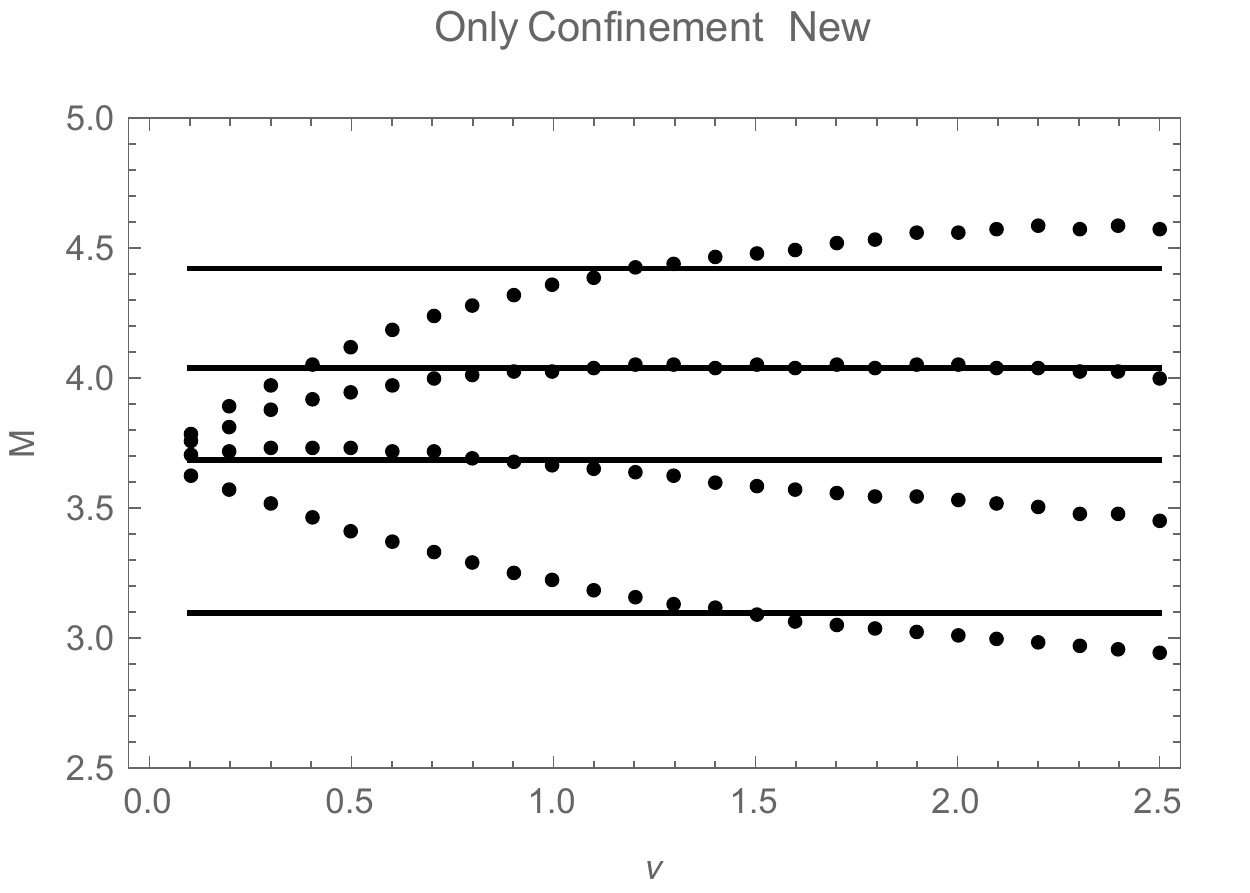}\\
    \includegraphics[width=2.3 in]{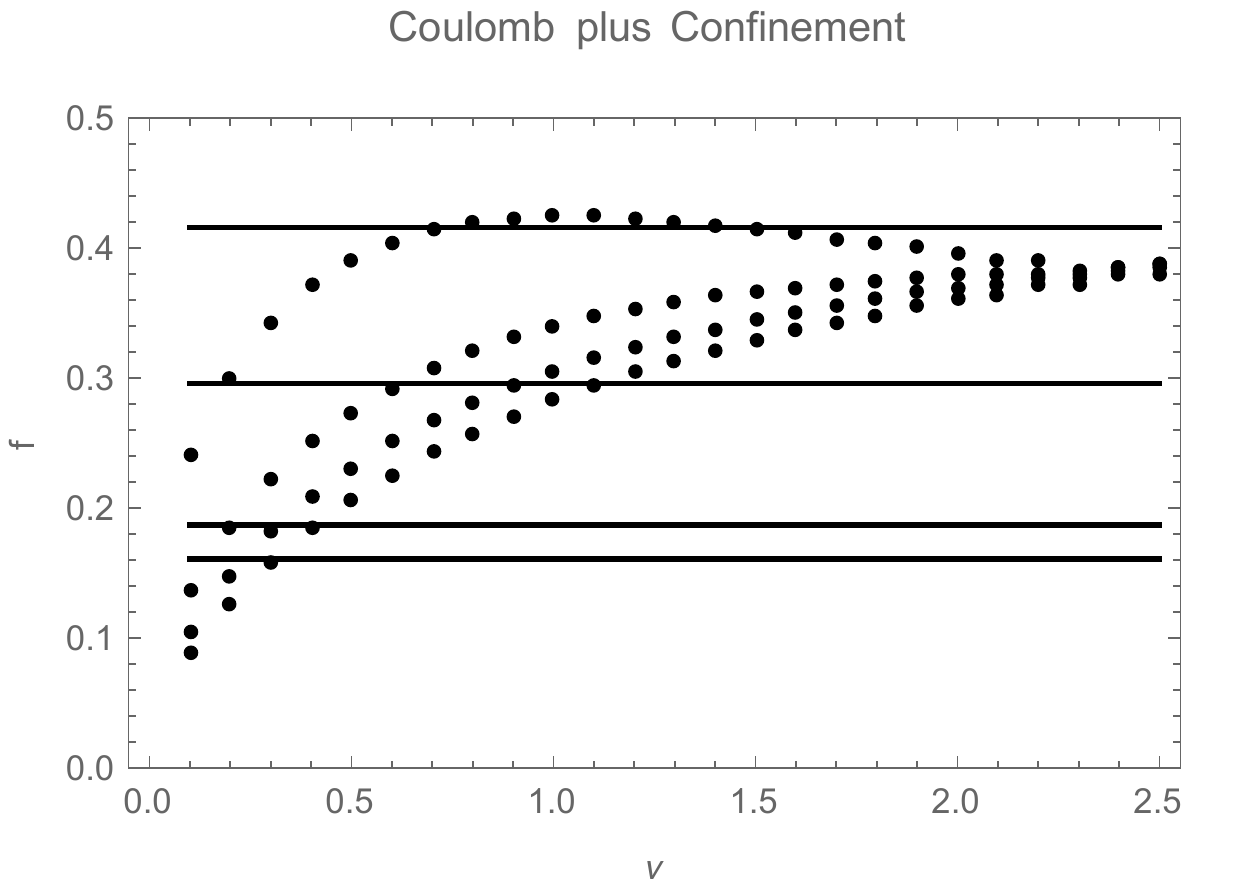}
    \includegraphics[width=2.3 in]{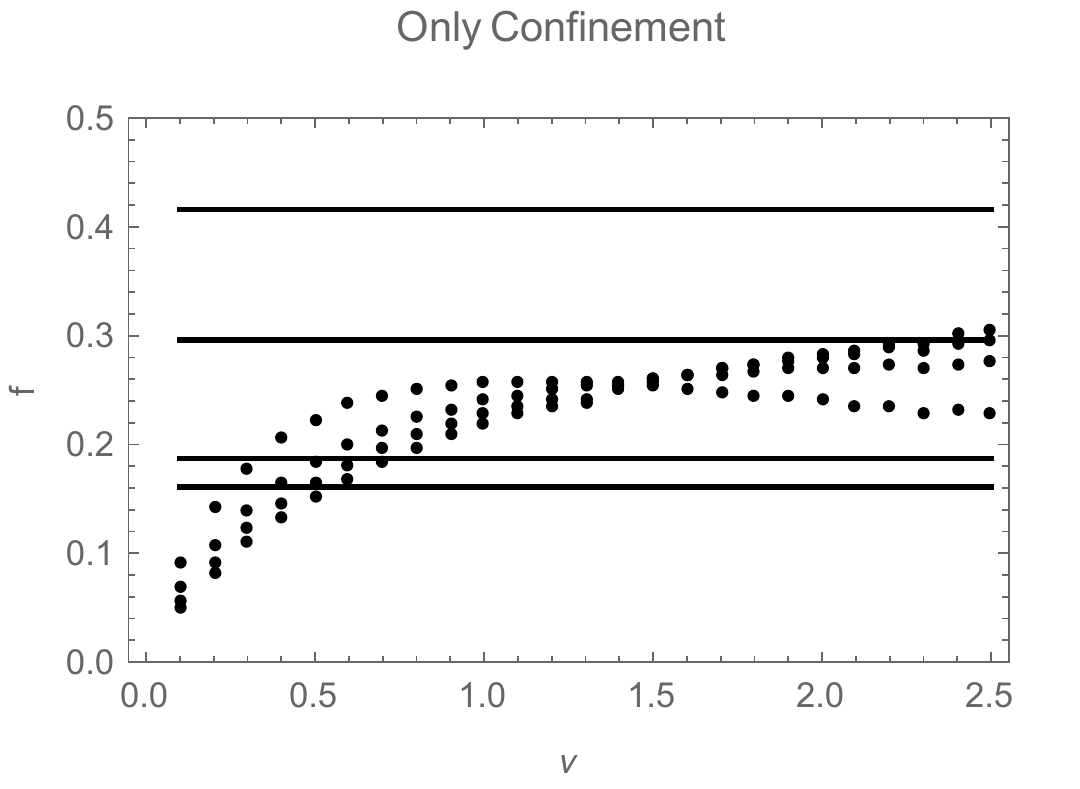}
    \includegraphics[width=2.3 in]{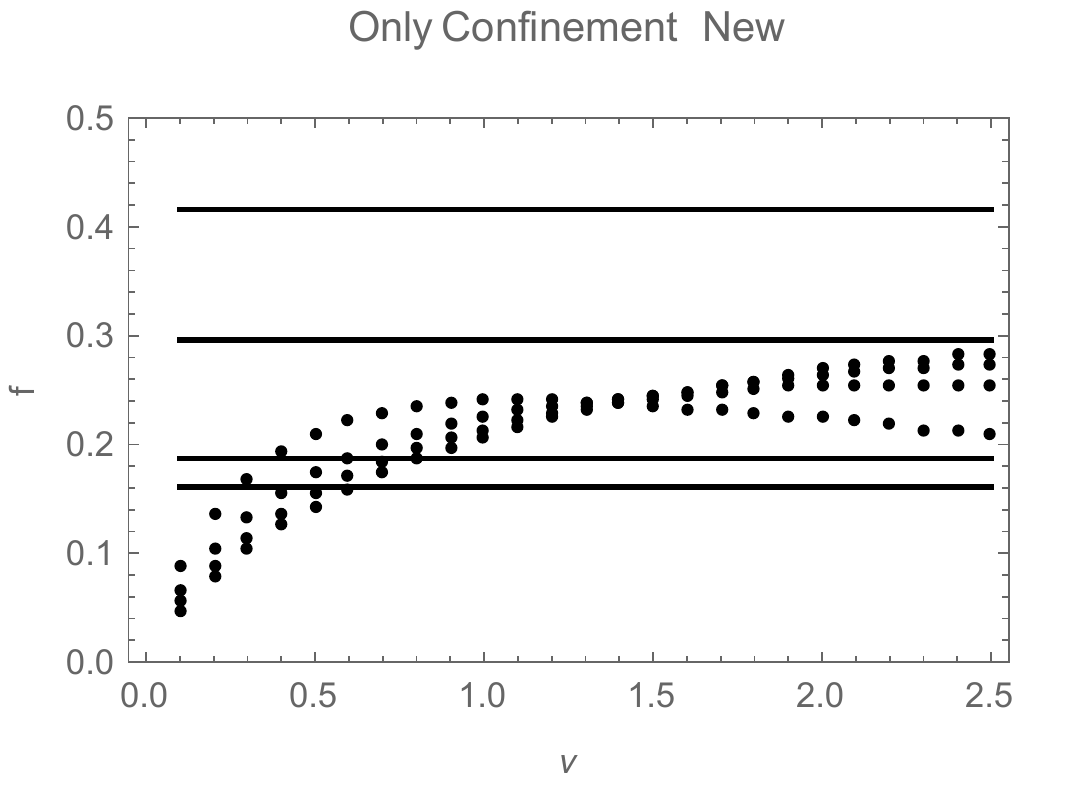}
  \end{tabular}
\caption{
Masses and decay constants of charmonium for the different values of $\nu$ considered in this work when $V_{0} = 0$. The continuous line corresponds to the experimental value of the masses for the considered states.}
\end{figure*}
\end{center}

\begin{table*}
\begin{center}
\begin{tabular}{| c | c | c | c | c | c | c |  c | c | c |}
\hline
$\nu$ & \multicolumn{3}{ c |}{$CC$} & \multicolumn{3}{ c |}{$OC$} & \multicolumn{3}{ c |}{$OCR$} \\ 
\cline{2-10}
& $A$ & $V_{0}$ & Error & $A$ & $V_{0}$ & Error & $A$ & $V_{0}$ & Error \\
\hline
\hline
~~0.1~~ & ~~5.81~~ & ~~5.78~~ & ~~0.020~~ & ~~5.81~~ & ~~5.78~~ & ~~0.075~~ & ~~7.14~~ & ~~7.47~~ & ~~0.015~~ \\ 
 \hline
~~0.2~~ & ~~2.59~~ & ~~2.55~~ & ~~0.017~~ & ~~2.59~~ & ~~2.55~~ & ~~0.071~~ & ~~3.14~~ & ~~3.44~~ & ~~0.013~~ \\
 \hline
~~0.3~~ & ~~1.50~~ & ~~1.43~~ & ~~0.015~~ & ~~1.50~~ & ~~1.43~~ & ~~0.070~~ & ~~1.80~~ & ~~2.05~~ & ~~0.011~~ \\ 
 \hline
~~0.4~~ & ~~0.99~~ & ~~0.90~~ & ~~0.012~~ & ~~0.99~~ & ~~0.90~~ & ~~0.069~~ & ~~1.18~~ & ~~1.40~~ & ~~0.009~~ \\ 
 \hline
~~0.5~~ & ~~0.68~~ & ~~0.56~~ & ~~0.010~~ & ~~0.68~~ & ~~0.56~~ & ~~0.067~~ & ~~0.81~~ & ~~0.99~~ & ~~0.008~~ \\
 \hline
~~0.6~~ & ~~0.49~~ & ~~0.34~~ & ~~0.009~~ & ~~0.49~~ & ~~0.34~~ & ~~0.068~~ & ~~0.58~~ & ~~0.72~~ & ~~0.008~~ \\ 
 \hline
~~0.7~~ & ~~0.36~~ & ~~0.19~~ & ~~0.008~~ & ~~0.36~~ & ~~0.19~~ & ~~0.064~~ & ~~0.43~~ & ~~0.54~~ & ~~0.008~~ \\
 \hline
~~0.8~~ & ~~0.27~~ & ~~0.07~~ & ~~0.007~~ & ~~0.27~~ & ~~0.07~~ & ~~0.066~~ & ~~0.32~~ & ~~0.40~~ & ~~0.009~~ \\ 
 \hline
~~0.9~~ & ~~0.21~~ & ~~0.00~~ & ~~0.008~~ & ~~0.21~~ & ~~0.00~~ & ~~0.065~~ & ~~0.24~~ & ~~0.28~~ & ~~0.011~~ \\
 \hline
~~1.0~~ & ~~0.16~~ & ~~-0.07~~ & ~~0.009~~ & ~~0.16~~ & ~~-0.07~~ & ~~0.062~~ & ~~0.18~~ & ~~0.18~~ & ~~0.012~~ \\ 
 \hline
~~1.1~~ & ~~0.12~~ & ~~-0.15~~ & ~~0.010~~ & ~~0.12~~ & ~~-0.15~~ & ~~0.063~~ & ~~0.14~~ & ~~0.11~~ & ~~0.013~~ \\
 \hline
~~1.2~~ & ~~0.10~~ & ~~-0.16~~ & ~~0.012~~ & ~~0.10~~ & ~~-0.16~~ & ~~0.064~~ & ~~0.11~~ & ~~0.07~~ & ~~0.014~~ \\ 
 \hline
~~1.3~~ & ~~0.07~~ & ~~-0.26~~ & ~~0.014~~ & ~~0.07~~ & ~~-0.26~~ & ~~0.064~~ & ~~0.09~~ & ~~0.05~~ & ~~0.017~~\\
 \hline
~~1.4~~ & ~~0.06~~ & ~~-0.25~~ & ~~0.014~~ & ~~0.06~~ & ~~-0.25~~ & ~~0.064~~ & ~~0.07~~ & ~~0.00~~ & ~~0.018~~ \\ 
 \hline
~~1.5~~ & ~~0.05~~ & ~~-0.25~~ & ~~0.017~~ & ~~0.05~~ & ~~-0.25~~ & ~~0.063~~ & ~~0.05~~ & ~~-0.09~~ & ~~0.019~~ \\
 \hline
~~1.6~~ & ~~0.04~~ & ~~-0.27~~ & ~~0.019~~ & ~~0.04~~ & ~~-0.27~~ & ~~0.063~~ & ~~0.04~~ & ~-0.11~~ & ~~0.020~~ \\ 
 \hline
~~1.7~~ & ~~0.03~~ & ~~-0.32~~ & ~~0.018~~ & ~~0.03~~ & ~~-0.32~~ & ~~0.063~~ & ~~0.03~~ & ~~-0.16~~ & ~~0.022~~ \\
 \hline
~~1.8~~ & ~~0.02~~ & ~~-0.41~~ & ~~0.022~~ & ~~0.02~~ & ~~-0.41~~ & ~~0.065~~ & ~~0.03~~ & ~~-0.07~~ & ~~0.027~~ \\ 
 \hline
~~1.9~~ & ~~0.02~~ & ~~-0.33~~ & ~~0.022~~ & ~~0.02~~ & ~~-0.33~~ & ~~0.062~~ & ~~0.02~~ & ~~-0.18~~ & ~~0.022~~ \\
 \hline
~~2.0~~ & ~~0.01~~ & ~~-0.51~~ & ~~0.037~~ & ~~0.01~~ & ~~-0.51~~ & ~~0.069~~ & ~~0.02~~ & ~~-0.09~~ & ~~0.032~~ \\ 
 \hline
~~2.1~~ & ~~0.01~~ & ~~-0.45~~ & ~~0.025~~ & ~~0.01~~ & ~~-0.45~~ & ~~0.064~~ & ~~0.01~~ & ~~-0.31~~ & ~~0.032~~ \\
 \hline
~~2.2~~ & ~~0.01~~ & ~~-0.39~~ & ~~0.024~~ & ~~0.01~~ & ~~-0.39~~ & ~~0.065~~ & ~~0.01~~ & ~~-0.23~~ & ~~0.025~~ \\ 
 \hline
~~2.3~~ & ~~0.01~~ & ~~-0.32~~ & ~~0.037~~ & ~~0.01~~ & ~~-0.32~~ & ~~0.070~~ & ~~0.01~~ & ~~-0.16~~ & ~~0.033~~ \\
 \hline
~~2.4~~ & ~~0.01~~ & ~~-0.24~~ & ~~0.056~~ & ~~0.01~~ & ~~-0.24~~ & ~~0.080~~ & ~~0.01~~ & ~~-0.08~~ & ~~0.050~~ \\ 
 \hline
~~2.5~~ & ~~0.01~~ & ~~-0.17~~ & ~~0.078~~ & ~~0.01~~ & ~~-0.17~~ & ~~0.098~~ & ~~0.01~~ & ~~0.01~~ & ~~0.071~~ \\
 \hline
 \hline
\end{tabular}
\caption{Parameters of the potential for the charmonium next to the corresponding rms error obtained for the masses of the four charmonia considered in this work for the case where $V_ {0}\neq 0$. The $CC$ and $OC$ indices consider that these parameters were obtained considering a Coulomb potential plus confinement ($CC$), confinement only ($OC$) or confinement only with a readjustment of parameters ($OCR$).}
\label{tabla1}
\end{center}
\end{table*}

\begin{center}
\begin{figure*}
  \begin{tabular}{c c c}
    \includegraphics[width=2.3 in]{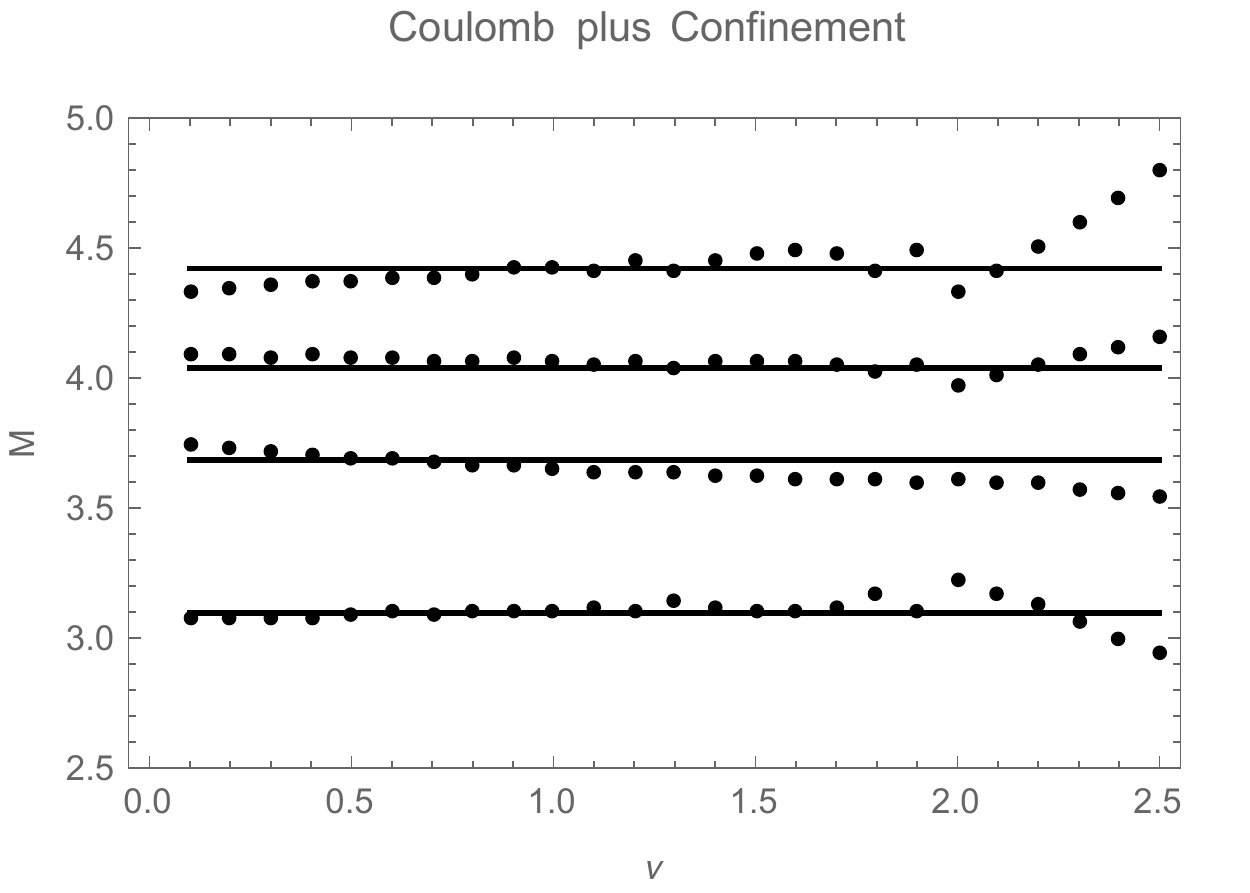}
    \includegraphics[width=2.3 in]{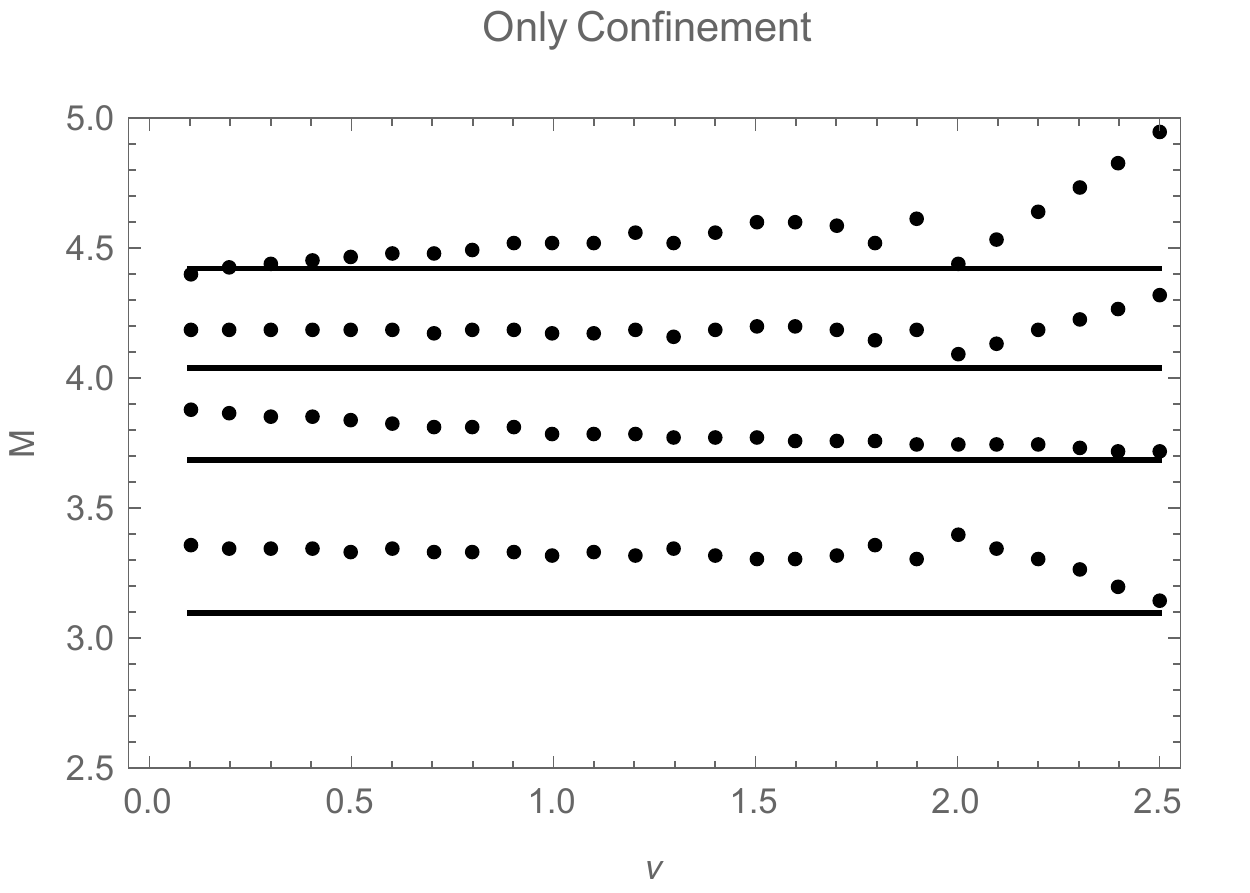}
    \includegraphics[width=2.3 in]{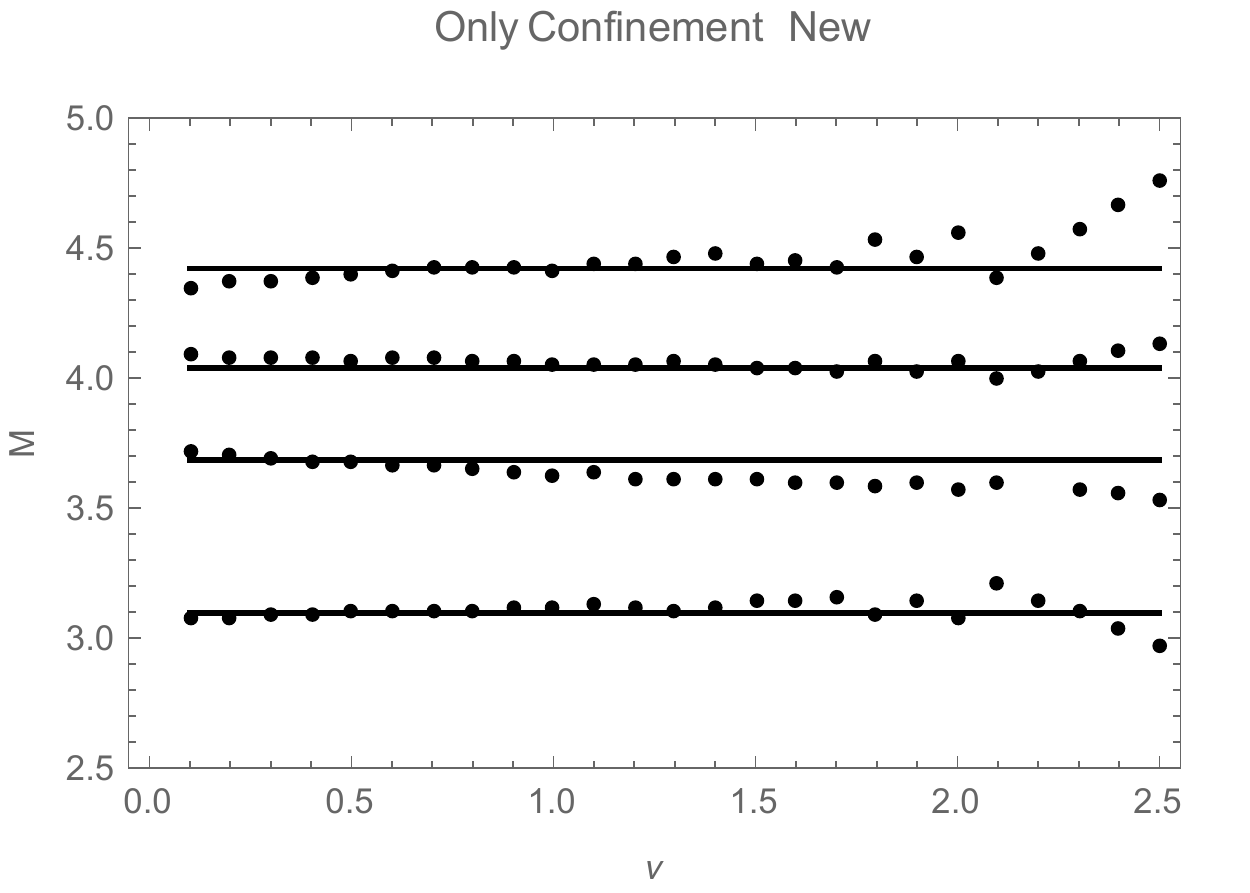}\\
    \includegraphics[width=2.3 in]{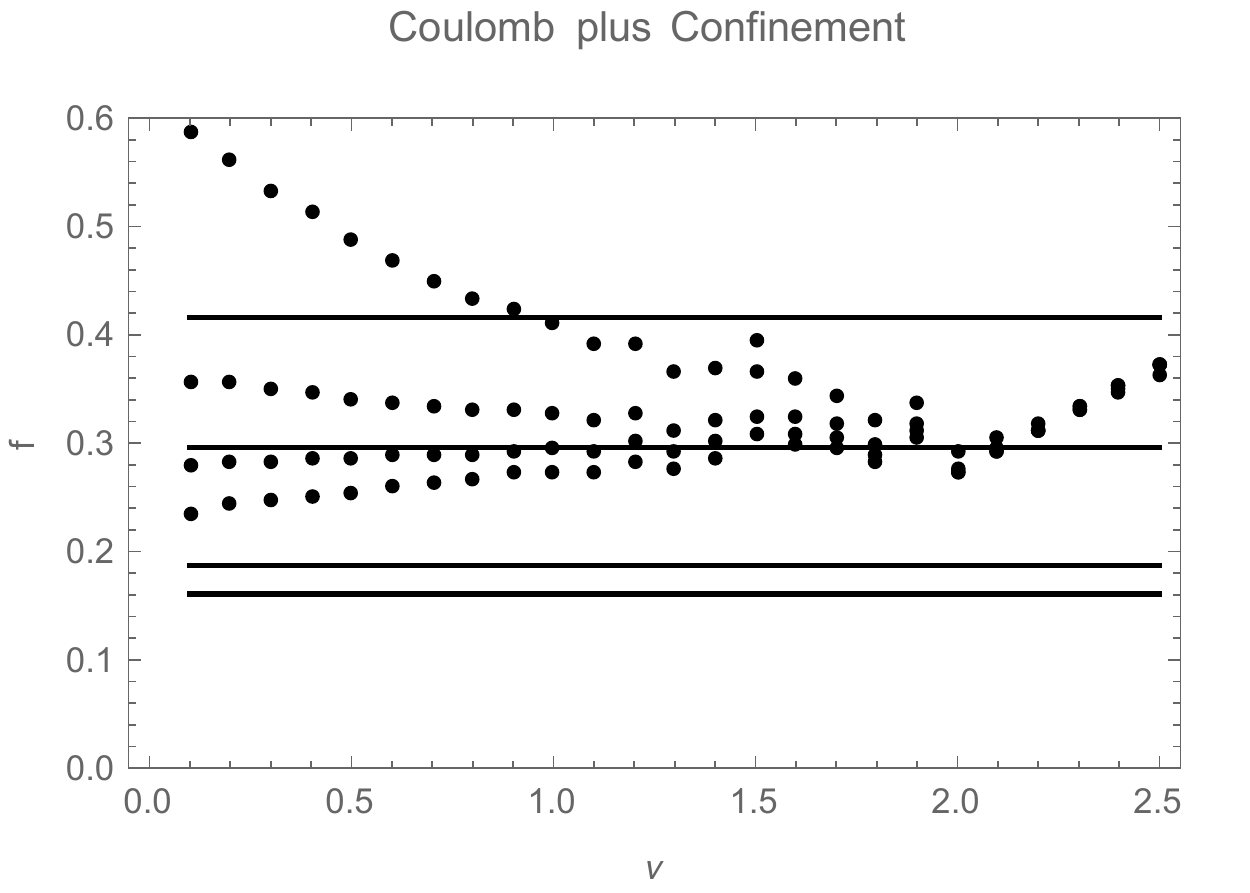}
    \includegraphics[width=2.3 in]{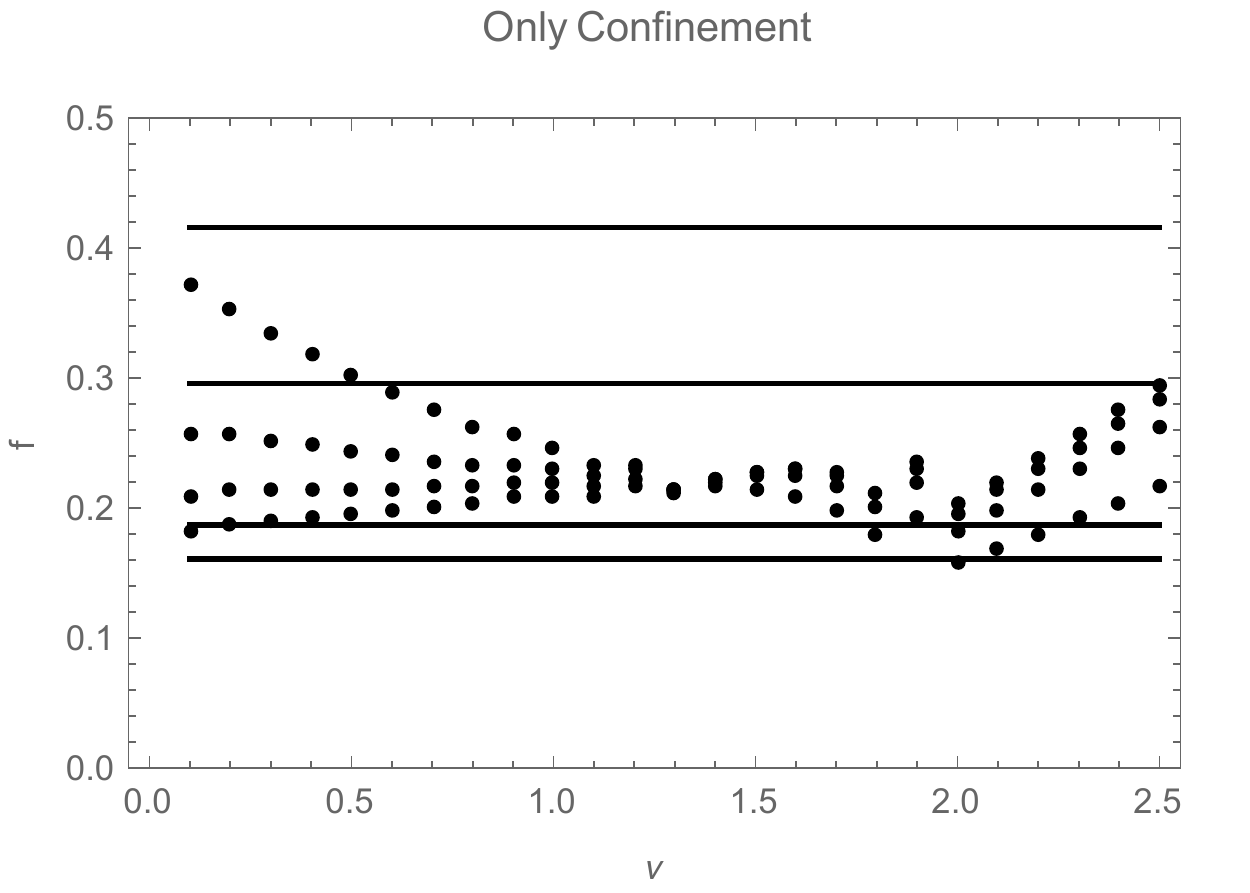}
    \includegraphics[width=2.3 in]{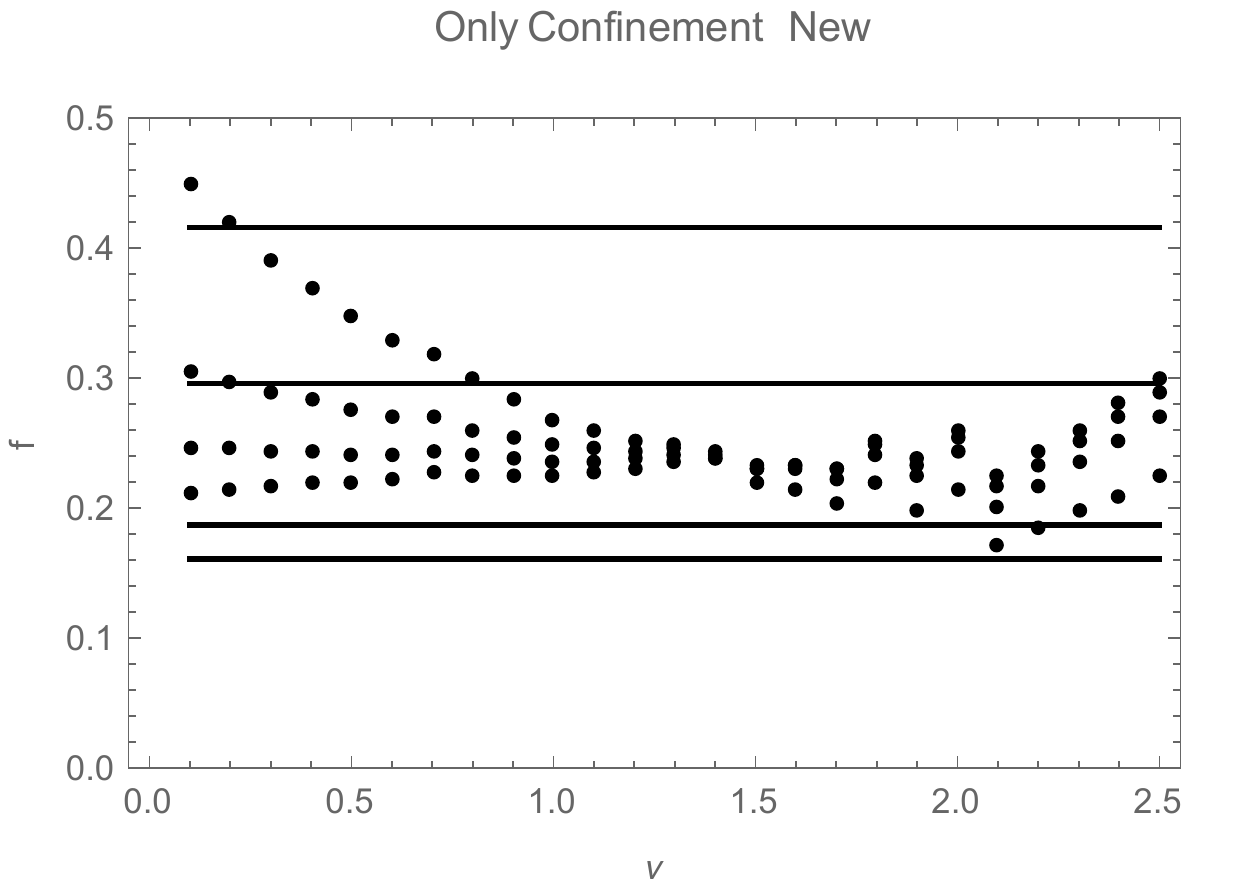}
  \end{tabular}
\caption{Masses and decay constants of charmonia for the different values of $\nu$ considered in this work when $V_{0}$ is adjusted. The continuous line corresponds to the experimental value of the masses for the considered states.}
\end{figure*}
\end{center}




\section{Analysis of the dominant term in the potential}

One of the aspects that we are interested in studying is when the confinement part is larger than the Coulomb part in the full potential. For this purpose we use for each $\nu$ the criterion discussed in \cite{Choudhury:2013kta, Choudhury:2014mva} to know when the linear confinement potential dominates the Coulomb part in the Cornell potential, i.e., for each $\nu$ we look for the $r_{0}$ that satisfies $-\frac{\alpha_{c}}{r_{0}} + A r_{0}^{\nu} = 0$, which gives us a value that marks the transition in which the Coulomb term is greater than the confinement term in the potential ($r < r_{0}$) or when the confinement term is greater ($r > r_{0}$).




Then we calculate the average $r$ for the state $1S$ ($\langle r \rangle_ {1S}$), which corresponds to the average radius of the more compact charmonium, and we compare it with the $r_{0}$ mentioned above, so we can get an idea of whether, for each value of $\nu$, in charmonium the confinement term is greater or not than the Coulomb contribution.


Table III shows the values of $r_{0}$ and $\langle r \rangle_{1S}$ for the cases considered, and FIG. 3 shows these values for each value of $\nu$.


\begin{table}
\begin{center}
\begin{tabular}{| c | c | c | c | c |}
\hline
$\nu$ & \multicolumn{2}{ c |}{$V_{0} = 0$} & \multicolumn{2}{ c |}{$V_{0} \neq 0$}  \\ 
\cline{2-5}
& $r_{0}$ & $\langle r \rangle_{1S}$ & $r_{0}$ & $\langle r \rangle_{1S}$ \\
\hline
  
 \hline
 ~~0.1~~ & ~~0.49~~ & ~~3.44~~& ~~0.09~~& ~~1.82~~ \\ 
 \hline
 ~~0.2~~ & ~~0.60~~ & ~~2.87~~& ~~0.21~~& ~~1.86~~ \\
 \hline
 ~~0.3~~ & ~~0.72~~ & ~~2.59~~& ~~0.36~~& ~~1.91~~ \\
 \hline
 ~~0.4~~ & ~~0.83~~ & ~~2.42~~& ~~0.52~~& ~~1.95~~ \\
 \hline
 ~~0.5~~ & ~~0.95~~ & ~~2.32~~& ~~0.70~~& ~~2.01~~ \\
 \hline
 ~~0.6~~ & ~~1.07~~ & ~~2.26~~& ~~0.88~~& ~~2.05~~ \\
 \hline
 ~~0.7~~ & ~~1.18~~ & ~~2.21~~& ~~1.06~~& ~~2.10~~ \\
 \hline
 ~~0.8~~ & ~~1.30~~ & ~~2.19~~& ~~1.24~~& ~~2.14~~ \\
 \hline
 ~~0.9~~ & ~~1.41~~ & ~~2.17~~& ~~1.40~~& ~~2.17~~ \\
 \hline
 ~~1.0~~ & ~~1.52~~ & ~~2.17~~& ~~1.58~~& ~~2.22~~ \\
 \hline
 ~~1.1~~ & ~~1.62~~ & ~~2.16~~& ~~1.77~~& ~~2.28~~ \\
 \hline
 ~~1.2~~ & ~~1.73~~ & ~~2.17~~& ~~1.88~~& ~~2.27~~ \\
 \hline
 ~~1.3~~ & ~~1.83~~ & ~~2.17~~& ~~2.13~~& ~~2.38~~ \\
 \hline
 ~~1.4~~ & ~~1.93~~ & ~~2.18~~& ~~2.20~~& ~~2.36~~ \\
 \hline
 ~~1.5~~ & ~~2.02~~ & ~~2.19~~& ~~2.30~~& ~~2.37~~ \\
 \hline
 ~~1.6~~ & ~~2.12~~ & ~~2.21~~& ~~2.42~~& ~~2.40~~ \\
 \hline
 ~~1.7~~ & ~~2.21~~ & ~~2.22~~& ~~2.61~~& ~~2.46~~ \\
 \hline
 ~~1.8~~ & ~~2.30~~ & ~~2.24~~& ~~2.92~~& ~~2.59~~ \\
 \hline
 ~~1.9~~ & ~~2.36~~ & ~~2.24~~& ~~2.81~~& ~~2.49~~ \\
 \hline
 ~~2.0~~ & ~~2.46~~ & ~~2.26~~& ~~3.42~~& ~~2.77~~ \\
 \hline
 ~~2.1~~ & ~~2.55~~ & ~~2.29~~& ~~3.29~~& ~~2.68~~ \\
 \hline
 ~~2.2~~ & ~~2.59~~ & ~~2.28~~& ~~3.17~~& ~~2.59~~ \\
 \hline
 ~~2.3~~ & ~~2.70~~ & ~~2.32~~& ~~3.06~~& ~~2.51~~ \\
 \hline
 ~~2.4~~ & ~~2.74~~ & ~~2.31~~& ~~2.96~~& ~~2.43~~ \\
 \hline
 ~~2.5~~ & ~~2.79~~ & ~~2.32~~& ~~2.87~~& ~~2.36~~ \\

 \hline
 \hline
\end{tabular}
\caption{$r_{0}$ and $\langle r \rangle_{1S}$ values for quarkonia considered in this work.}
\label{tabla1}
\end{center}
\end{table}

\begin{center}
\begin{figure*}
  \begin{tabular}{c c c}
    \includegraphics[width=3.4 in]{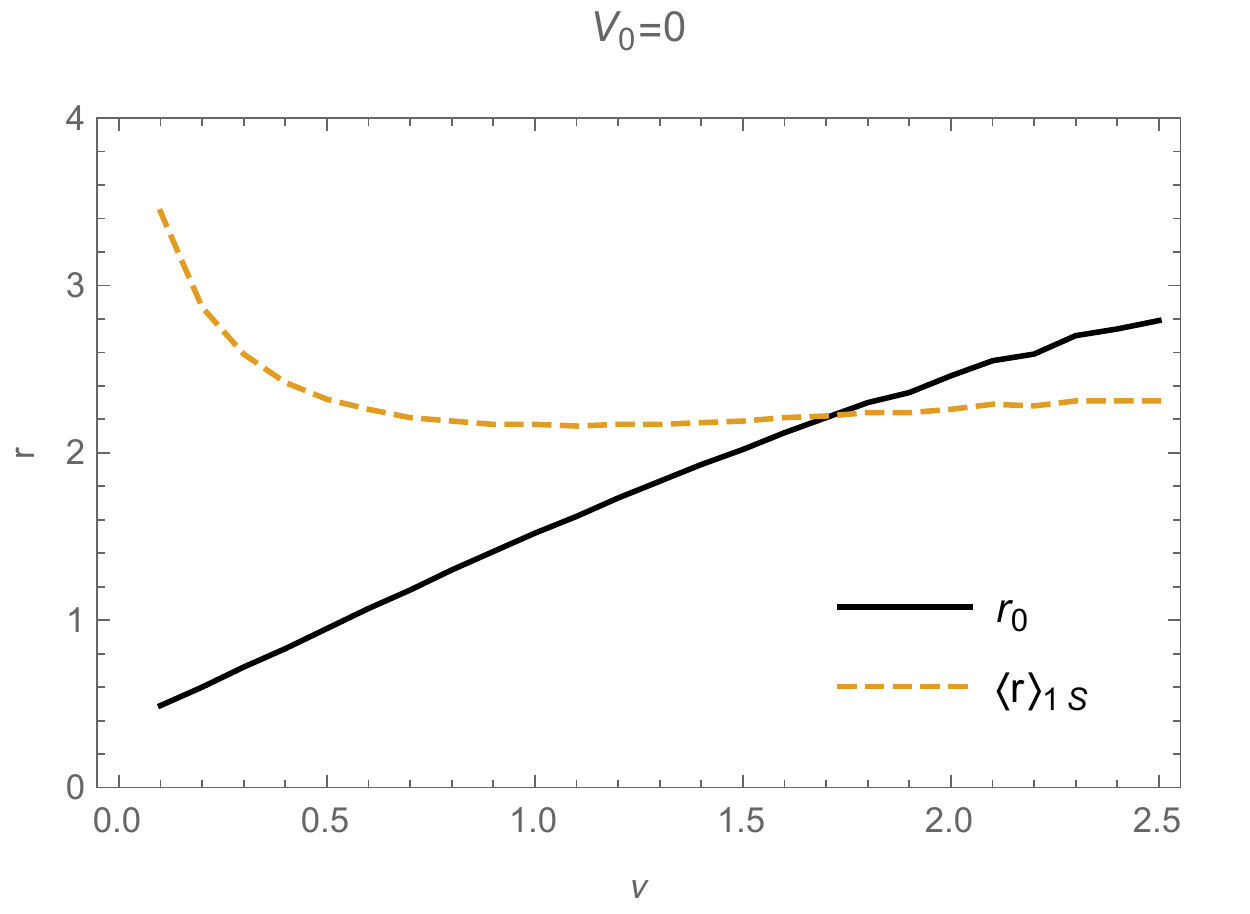}
    \includegraphics[width=3.4 in]{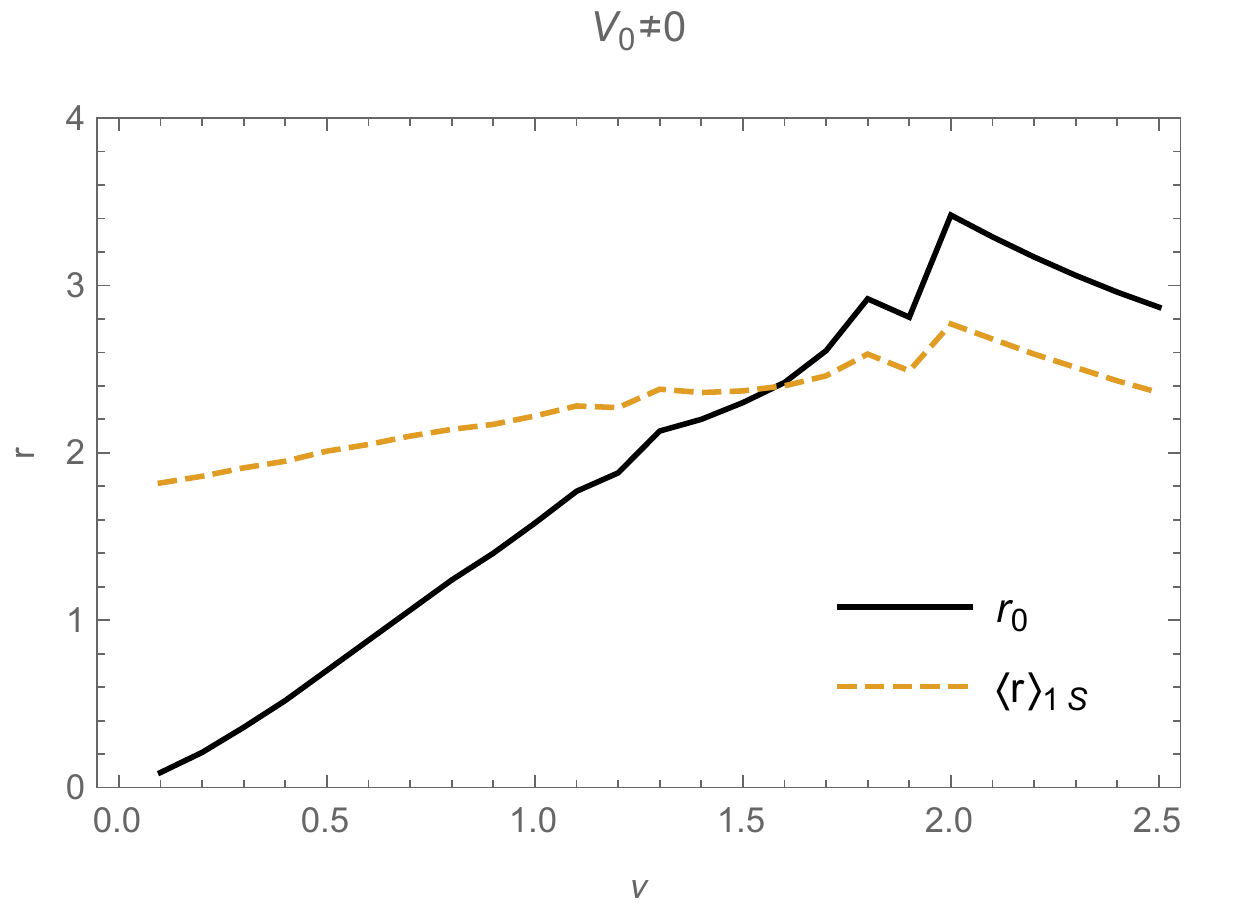}
  \end{tabular}
\caption{Comparison between $r_{0}$ and $\langle r\rangle_{1S}$ for each value of $\nu$ in cases $V_{0} = 0$ and $V_{0} \neq 0$, of according to the results of table III.}
\end{figure*}
\end{center}




\section{Discussion of results}

The main objective of this work was to analyze if heavy quarkonia can be studied using only confining potentials, understanding these as the dominant part of a potential well motivated by QCD that contains a Coulomb contribution due to the exchange of a gluon plus a confinement part. For this we have considered the "Coulomb plus power potential ($CPP\nu$)" used in \cite{Rai:2008sc, Patel:2008na}, which is of the form $V(r) = - \frac{\alpha_{c}}{r} + A r^{\nu} - V_{0}$, where on this occasion we have considered values of $\nu$ ranging from $0.1$ to $2.5$ in steps of $0.1$, and the parameters were determined considering on the one hand that $V_{0} = 0$ and that the only free parameter was $A$, and on the other that both $V_{0}$ and $A$ should be determined in order to obtain a good fit for the masses of the charmonium. In both cases, it was studied when the confinement term was greater than the contribution of the Coulomb part, a value that was compared to the average radius of the base state, which was suggested in \cite{Choudhury:2013kta, Choudhury:2014mva} for Cornell's potential it gives us an idea of when the confinement dominates over the Coulomb part. Figure 3 and Table III show the results for $r_{0}$ and $\langle r \rangle_{1S}$, and show that for the chosen set of parameters  the confinement term delivers a contribution equal to or greater than that of the part of Coulomb when $\nu <1.5$, both for the case $V_{0} = 0$ and $V_{0} \neq 0$.

We have calculated the masses and constants of decay of $J/\Psi$ and its first radial excitations considering the following cases: on the one hand the full potential (called $CC$), neglecting the Coulomb part without varying $A$ ($OC$) and determining again the value of $A$ ($OCR$). Figures 1 and 2 and Tables I and II show the results obtained in each case, and show that for the case $V_{0} = 0$ the best masses are obtained when $\nu \sim 1.0$, and for the case $V_{0} \neq 0$ a good description of the spectrum is possible for values of $\nu$ ranging between $0.1$ and $2.0$, while the results for the decay constants are poor in both cases, although it can be seen that for low values of $\nu$ when $V_{0}\neq0$ the simultaneous description of the observables that interest us in this work is better.


It is important to remember that at present it is known that a potential that aspires to describe mesons with heavy quarks appropriately must include corrections to the dominant part that we have considered here, because by introducing interactions between the spin of the quarks it is possible to differentiate between the spectrum of scalar and vector mesons. Moreover,such corrections modify the behavior of the wave function at the origin, which allows the calculations of decay constants to be significantly improved, while the mass spectrum undergoes changes that do not move as far from the spectrum obtained with the dominant contribution that the $CPP\nu$ model delivers.


From the above, particularly if we restrict ourselves to the results for the mass spectrum, we can see that for the case when $V_{0} \neq 0$, the potentials that only consider power law type confinement with indices less than $1.5$ can be considered as the dominant contribution in a well-motivated potential of QCD, i.e., potentials that consider contributions from the exchange of a gluon plus confinement, and shed light on why potentials like Martin's are successful at describing properties of mesons formed by heavy quarks.


\vspace{0.2cm}
\noindent
\textbf{Acknowledgment: } Work supported by FONDECYT (Chile) under Grant No. 1180753.


\begin{thebibliography}{99}

\bibitem{Aoki:2013ldr} 
  S.~Aoki {\it et al.},
  Eur.\ Phys.\ J.\ C {\bf 74}, 2890 (2014)
  doi:10.1140/epjc/s10052-014-2890-7
  [arXiv:1310.8555 [hep-lat]].

\bibitem{Roberts:1994dr} 
  C.~D.~Roberts and A.~G.~Williams,
  Prog.\ Part.\ Nucl.\ Phys.\  {\bf 33}, 477 (1994)
  doi:10.1016/0146-6410(94)90049-3
  [hep-ph/9403224].
  
\bibitem{Lucha:1991vn} 
  W.~Lucha, F.~F.~Schoberl and D.~Gromes,
  Phys.\ Rept.\  {\bf 200}, 127 (1991).
  doi:10.1016/0370-1573(91)90001-3
  
\bibitem{Bykov:1985it} 
  A.~A.~Bykov, I.~M.~Dremin and A.~V.~Leonidov,
  Sov.\ Phys.\ Usp.\  {\bf 27}, 321 (1984)
  [Usp.\ Fiz.\ Nauk {\bf 143}, 3 (1984)].
  doi:10.1070/PU1984v027n05ABEH004291
  
\bibitem{Dib:2012vw} 
  C.~O.~Dib and N.~Neill,
  Phys.\ Rev.\ D {\bf 86}, 094011 (2012)
  doi:10.1103/PhysRevD.86.094011
  [arXiv:1208.2186 [hep-ph]].
  
\bibitem{Rai:2008sc} 
  A.~K.~Rai, B.~Patel and P.~C.~Vinodkumar,
  Phys.\ Rev.\ C {\bf 78}, 055202 (2008)
  doi:10.1103/PhysRevC.78.055202
  [arXiv:0810.1832 [hep-ph]].
  
\bibitem{Patel:2008na} 
  B.~Patel and P.~C.~Vinodkumar,
  J.\ Phys.\ G {\bf 36}, 035003 (2009)
  doi:10.1088/0954-3899/36/3/035003
  [arXiv:0808.2888 [hep-ph]].
  
\bibitem{Martin:1980jx} 
  A.~Martin,
  Phys.\ Lett.\  {\bf 93B}, 338 (1980).
  doi:10.1016/0370-2693(80)90527-4
  
\bibitem{Martin:1980rm} 
  A.~Martin,
  Phys.\ Lett.\  {\bf 100B}, 511 (1981).
  doi:10.1016/0370-2693(81)90617-1
  
\bibitem{Eichten:1974af} 
  E.~Eichten, K.~Gottfried, T.~Kinoshita, J.~B.~Kogut, K.~D.~Lane and T.~M.~Yan,
  Phys.\ Rev.\ Lett.\  {\bf 34}, 369 (1975)
  Erratum: [Phys.\ Rev.\ Lett.\  {\bf 36}, 1276 (1976)].
  doi:10.1103/PhysRevLett.34.369, 10.1103/PhysRevLett.36.1276
  
\bibitem{Eichten:1978tg} 
  E.~Eichten, K.~Gottfried, T.~Kinoshita, K.~D.~Lane and T.~M.~Yan,
  Phys.\ Rev.\ D {\bf 17}, 3090 (1978)
  Erratum: [Phys.\ Rev.\ D {\bf 21}, 313 (1980)].
  doi:10.1103/PhysRevD.17.3090, 10.1103/physrevd.21.313.2
  
\bibitem{Eichten:1979ms} 
  E.~Eichten, K.~Gottfried, T.~Kinoshita, K.~D.~Lane and T.~M.~Yan,
  Phys.\ Rev.\ D {\bf 21}, 203 (1980).
  doi:10.1103/PhysRevD.21.203

\bibitem{Choudhury:2013kta} 
  D.~K.~Choudhury and K.~K.~Pathak,
  arXiv:1304.7074 [hep-ph].
  
\bibitem{Choudhury:2014mva} 
  D.~K.~Choudhury and K.~K.~Pathak,
  J.\ Phys.\ Conf.\ Ser.\  {\bf 481}, 012022 (2014).
  doi:10.1088/1742-6596/481/1/012022
  
\bibitem{Lucha:1998xc} 
  W.~Lucha and F.~F.~Schoberl,
  Int.\ J.\ Mod.\ Phys.\ C {\bf 10}, 607 (1999)
  doi:10.1142/S0129183199000450
  [hep-ph/9811453].
  
\end{thebibliography}
\end{document}